\documentclass[12pt,a4paper]{article}
\makeatletter
\usepackage{rotating}
\usepackage{float}
\usepackage{siunitx}
\usepackage{mathrsfs}  
\usepackage[latin1]{inputenc}
\usepackage{amsmath}
\usepackage{amssymb}
\usepackage{amsfonts}
\usepackage{graphicx}
\usepackage{color}
\usepackage[numbers]{natbib}
\usepackage{array} 
\usepackage{mathrsfs}  
\usepackage{url}  
\usepackage{natbib}
 \usepackage{mathabx}
\usepackage{pgf,tikz}
\usepackage{booktabs} 
\usepackage{colortbl} 
\usepackage{xcolor} 
\usepackage{xfrac}
\usepackage{makecell}
\usetikzlibrary{cd}
\newcommand{\blind}{1}
\newcommand{\indep}{\perp\mkern-8.mu\perp}
\newcommand{\noindep}{\phantom{\perp} /\mkern-18.mu\indep}

\begin{document}

\def\spacingset#1{\renewcommand{\baselinestretch}%
{#1}\small\normalsize} \spacingset{1}
\date{ }


\if1\blind
{
\title{\bf Estimating populational-average hazard ratios in the presence of unmeasured confounding} 
 \author{Pablo Mart\'inez-Camblor$^{1,2,\,}$\thanks{{\it Correspondence to:} Pablo Mart\'inez-Camblor. 7 Lebanon Street, Suite 309, Hinman Box 7261, Lebanon, NH 03751, USA. E-mail: {\color{blue} Pablo.Martinez-Camblor@hitchcock.org}},\\ 
 Todd A. MacKenzie$^{2,3}$, and A. James O'Malley$^{3,2}$
\hspace{.2cm}\\
 \small{$^1$Department of Anesthesiology, Dartmouth-Hitchcock Medical Center, NH, USA}\\ 
 \small{$^2$Department of Biomedical Data Science, Geisel School of Medicine at Dartmouth, NH, USA}\\
 \small{$^3$The Dartmouth Institute for Health Policy and Clinical Practice, NH, USA}}
  \maketitle
} \fi

\begin{abstract}
The Cox regression model and its associated hazard ratio (HR) are frequently used for summarizing the effect of treatments on
time to event outcomes. However, the HR's interpretation strongly depends on the assumed underlying survival model. The challenge of interpreting the HR has been the focus of a number of recent works. Besides, several alternative measures have been proposed in order to deal with these concerns. The marginal Cox regression models include an identifiable hazard ratio without individual but populational causal interpretation. In this work, we study the properties of one particular marginal Cox regression model and consider its estimation in the presence of omitted confounder. We prove the large sample consistency of an estimation score which allows non-binary treatments. Our Monte Carlo simulations suggest that finite sample behavior of the procedure is adequate. The studied estimator is more robust than its competitors for weak instruments although it is slightly more biased for large effects of the treatment. The practical use of the presented techniques is illustrated through a real practical example using data from the vascular quality initiative registry. The used R code is provided as Supplementary Material.

\end{abstract}
\noindent%
\vfill

\spacingset{1.45} 
\section{Introduction}

In biomedicine, proportional hazard Cox regressions \cite{cox72} are commonly used for modeling time-to-event outcomes. Besides, related hazard ratios (HR) are used for measuring the impact of the factors of interest on these outcomes. When the Cox regression assumptions hold, the HR represents the multiplicative effect
on the risk of having the studied event when the value of the characteristic increases by one unit and the rest of the variables involved in the survival function keep the same value. In absence of omitted covariates, assuming that the value of the studied variable, $X$, can be changed without changing the values of the rest of the covariates present in the problem, 
the HR is a {\it causal hazard ratio} associated with the studied event and measures the average change in the risk produced when $X$ increases one unit. One relevant handicap of Cox regression models is the loss of causal interpretation if a relevant covariate is not included in the model, even when there is no interaction between $X$ and this covariate \citep{aalen15, martinussen13, martinussen20}.

A conventional approach for addressing the unmeasured confounding is to consider the possibility of obtaining additional information through the so-called instrumental variable. These methodologies have been successfully implemented for dealing with endogeneity on standard regression models \citep{hernan06, angrist96, pearl00, tan06, wang18} including accelerated failure time (AFT) regressions \citep{robins91}. However, its application to risks-based regressions is more controversial because of the difficulties of these procedures for dealing with individual random-effects. That is, with covariates not related with $X$ but affecting the outcome. \citet{camblor17} proposed a two-stage algorithm (2SRI-F) for addressing the endogeneity in proportional hazard Cox regression models. The 2SRI-F asks strong assumptions of the unmeasured confounders and, in a first stage, re-builds these confounders by adding individual frailties \citep{wienke10} to the Cox regression performed in a second stage. Beyond the required assumptions, the main handicap of this procedure is the difficulties of computing, in practice, exact individual frailties \cite{camblor18}. \citet{todd14} proposed an alternative survival model in which the omitted covariates do not impact the studied treatment when they are drawn from independent random variables. Under this Cox-type model, they proposed an estimating equation based on the orthogonality of instrument with Martingale residuals. \citet{wang18b} considered binary treatments and binary instrumental variables for introducing a weighting function which, under several assumptions, allows derivation of an unbiased estimator for the populational-average hazard ratio. 

In this work, we are interested in the estimation of the populational-average hazard ratios in the presence of omitted confounders. We consider the same structural model used in \citet{wang18b} and prove that the estimation equation proposed by \citet{todd14} provides a consistent estimator for the targeted parameter. Unlike  \citet{wang18b}, the considered procedure allows the study of continuous treatment and the use of continuous instruments. Besides, it allows adjustment by potential measured confounders. 

The paper is organized as follows. In Section \ref{frame}, we establish the framework and introduce the counterfactual and counting processes notation and the assumptions subsequently required. The goal of the marginal structural Cox regression model is introduced in Section \ref{marginal}. We also prove that the standard estimator based on the maximization of the partial likelihood function provides a consistent estimator of the targeted populational hazard ratio when the study design is a randomized clinical trial (RCT) with perfect compliance. We also develop two different approaches for dealing with measured covariates. The case in which the failure time function can be affected by omitted covariates is considered in Section \ref{iv}. We prove the consistency of the proposed estimation procedure and its asymptotic distribution. The finite-sample behavior is studied through Monte Carlo simulations in Section \ref{mcs}. Results show the consistency of the studied procedure and the validity of the normal approximation in moderate sample-size scenarios. In Section \ref{example}, we  compare the behavior of patients undergoing transcarotid artery revascularization versus those undergoing transfemoral carotid artery stenting using data from the Vascular Quality Initiative registry (\url{www.vascularqualityinitiative.org}) from January 2016 to May 2020. Proofs of the results are relegated to the Appendix. The \url{R} code used for implementing the considered procedures in our real-world example are provided as Supplementary material.

\section{General framework and notation}
\label{frame}
We consider an experiment in which our interest is to know the impact of the variable $X$ on the failure time $T$. We assume that the levels of $X$ could be modified without changing the rest of the experiment conditions and the potential presence of a censoring time $C$ such that the observed time is given by $Z=\min(T,C)$. We also know the event indicator $\Delta$ ($= I(T\leq C)$). Given $U$ a vector of covariates, we have that the survival function of the time $T$ is
\begin{equation}
S_x(t|U=u) = {\mathscr P}\{ T > t| X=x, U=u\}= e^{-\Lambda_x(t;u)},
\end{equation}
where $\Lambda_x(t;u)=\int_0^t \lambda_x(s;u)ds$ is the cumulative hazard with $\lambda_x(s;u)$ being the hazard function. Let $\{(z_i, \delta_i , x_i, u_i)\}_{i=1}^n$ be an independent and identically distributed (iid) random sample from $(Z,\Delta , X, U)$. We suppose that the 
counting process, $N_i(t)$, which records the number of failures, $0$ or $1$, in $[0, t]$ for the $i$-th subject ($1\leq i\leq n$), is right continuous and that no two components have simultaneous jumps \citep{andersen82}. Let be the intensity process
\begin{equation}
{\cal I}_i(t) = Y_i(t)\cdot \lambda_{x_i}(t; u_i)\quad (i=1,\cdots , n),
\end{equation}
where $Y_i(t)$ is a predictable process, it takes the value $1$ if the $i$-th subject is at risk just before $t$ and $0$ otherwise. The standard proportional hazard Cox regression model  \citep{cox75} assumes that $\lambda_x(s;u)=\lambda_0(s)\cdot\exp\{\beta_X\cdot x + \beta_u\cdot u\}$, where $\lambda_0(\cdot)$ is the baseline risk, $\beta_X$ and $\beta_u$ are the targeted parameters. The partial likelihood estimator is the solution to the equation
\begin{equation}
{\mathscr U}^X_n(\beta)=\sum_{i=1}^n \int_0^{\infty}\left\{x_i - \frac{{\cal S}^{(1)}_n(X,\beta,s)}{{\cal S}^{(0)}_n(X,\beta,s)}\right\}dN_i(s)=0,\label{pl}
\end{equation} 
where ${\cal S}^{(j)}_n(X,\beta,s)=n^{-1}\cdot \sum_{i=1}^n x_i^j\cdot Y_i(s)\cdot \exp\{\beta\cdot x_i\}$ ($j\in\mathbb N$), which is an appealing estimator for $\beta_x$ when $\beta_u=0$.

The (individual) {\it causal hazard ratio} refers to the relationship between the actual observed failure time and the counterfactual (failure time) that each subject would have experienced if the treatment would have been different to what it actually was. Adopting the potential outcome scheme \citep{rubin74}, let $T(x,d)$ and $C(x,d)$ be the potential failure and censoring times if the exposure would have been $d$ when it actually was $x$, respectively. It is well-known that, if any covariate affecting the failure time is omitted, the solution to the Eq. \ref{pl} is not an estimator for the causal hazard ratio \citep{aalen15}. The immediate consequence is that, the log-causal hazard ratio of a treatment on a time-to-event outcome following a standard proportional hazard Cox regression model is not identifiable even when the treatment is exogenous \citep{camblor20}.

When the sample is collected from an observational design, an additional problem to the previously reported no-identifiability is the potential presence of selection bias. To deal with this endogeneity, we assume that we can measure the (instrumental) variable $W$ which satisfies,
\vskip 0.5 cm
{\bf Assumption 1}. $W\indep (T,C)|X$.\par
{\bf Assumption 2}. $W\indep (U|X)$.\par
{\bf Assumption 3}. $W\noindep X$.\\
\vskip 0.1 cm
Assumption 1 guarantees that both the potential failure and censoring times just depend on the actual treatment received.  Assumption 2 implies that the information provided by the instrumental variable is independent (and not mixed with) the omitted confounder. The addition of measured confounders into Assumptions 1, 2 and 3, is straight forward.

\section{The marginal structural Cox model}
\label{marginal}
The objective of the marginal structural Cox model \citep{hernan01} is to develop a Cox-type model whose parameter of interest is identifiable when the omitted covariate and the studied treatment are independent ($X\indep U$). With this goal, we consider that the marginal survival function follows a proportional hazards model. That is, it is a standard proportional hazards Cox regression model when the effect of the independent covariates is integrated out. Therefore, we define the independent integrated Cox (IIC) model through
\begin{equation}
S_x(\cdot)= S_0(\cdot)^{e^{\beta_X\cdot x}}. \label{mcox}
\end{equation}
where $S_0(\cdot)={\mathbb E}_U\{e^{-\Lambda_0(\cdot;u)}\}$. That is, it satisfies that
\begin{align*}
S_x(t)=& {\mathbb E}_U\{e^{-\Lambda_x(t;u)}\}={\mathbb E}_U\left\{{\mathscr P}\{T>t|X=x, U=u\} \right\}\nonumber \\
=&{\mathbb E}_U\left\{{\mathscr P}\{T>t|X=0, U=u\} \right\}^{e^{\beta_X\cdot x}}= {\mathbb E}_U\{e^{-\Lambda_0(t;u)}\}^{e^{\beta_X\cdot x}}= S_0(t)^{e^{\beta_X\cdot x}}, 
\end{align*}
Example of this model can be found, for instance, in \citet{todd14, wang18b, wang18} or \citet{todd21}, among others. The targeted parameter, $\text{HR}_X$ ($=\exp\{\beta_X\}$), could change with the underlying study population, $U$, and it represents the effect of the treatment in the population under study. In this sense, it allows computation of  the populational counterfactual failure time for this population. That is, it represents the ratio of the risks of experiencing the outcome event when the whole population is in the same treatment group compared to the risk if the whole population is in a treatment level one unit below. In other words, the individual causal hazard ratio considers each subject compared with themselves under different treatments while the populational hazard ratio considers the comparison of each subject in the population with each subject in the same population that has a different treatment. The IIC model does not allow to {\it adjust} for unmeasured covariates but guarantees that the comparison is based on populations with the same characteristics.

Remark that, the IIC model conditional on $U$ is not assumed to be a Cox model. Rather, we assume that this particular survival distribution, whose form is unknown, has the property that, after integrating over the distribution of $U$ (which is the distribution of $U|X$ under randomization), we obtain a Cox model only depending on $X$. 

Several designs, for instance the randomized clinical trials, respect this assumption that the omitted covariates and the treatment are independent ($U\indep X$), and so allow the use of standard survival curve estimators. The next result proves that, in this case, the solution to the Eq. \ref{pl} yields a consistent estimator for $\beta_X$.\par

\vskip 0.8cm
\noindent
{\bf Theorem 1.} Let $\{(z_i, \delta_i , x_i, u_i)\}_{i=1}^n$ be an iid random sample from $(Z,\Delta , X, U)$ where the covariate $U$ is independent of both the censor time, $C$, and the treatment $X$ ($U\indep (C,X)$). If the underlying failure time satisfies Eq. \ref{mcox} and for each $t>0$,
$\int_0^t\mathbb E_X\{{\cal S}^{(2)}_n(\beta,s)\}ds<\infty$, then the solution to the function ${\mathscr U}^X_n(\cdot)$ in Eq. \ref{pl} is a consistent estimator of $\beta_X$.\par
$\hfill{\Box}$\par
\vskip 0.5 cm
\noindent
The IIC model can include directly measured covariates. However, the user has to decide between two different possibilities. Let $Q$ be a measured covariate. If we are interested in comparing subjects with the same levels of $Q$, we can consider the {\it adjusted} model, 
\begin{align*}
S_x(t|Q=q)=& {\mathbb E}_U\{e^{-\Lambda_x(t;u,q)}\}={\mathbb E}_U\left\{{\mathscr P}\{T>t|X=x, U=u, Q=q\} \right\}\nonumber \\
=&{\mathbb E}_U\left\{{\mathscr P}\{T>t|X=0, U=u, Q=0\} \right\}^{e^{\beta_X\cdot x + \beta_Q\cdot q}}\\ 
=& {\mathbb E}_U\{e^{-\Lambda_0(t;u)}\}^{e^{\beta_X\cdot x+\beta_Q\cdot q}}= S_{0,0}(t)^{e^{\beta_X\cdot x+\beta_Q\cdot q}}.
\end{align*}
If $Q\indep (U|X)$ and $\beta_X$ is known (or can be approximated), $\beta_Q$ can be consistently estimated from the solution to
\begin{equation*}
{\mathscr U}_n^Q(\beta)=\sum_{i=1}^n \int_0^{\infty}\left\{q_i - \frac{{\cal S}^{(1)}_n(Q,\beta,s)}{{\cal S}^{(0)}_n(Q,\beta,s)}\right\}dN_i(s)=0,\label{adj}
\end{equation*} 
where ${\cal S}^{(j)}_n(Q,\beta,s)=n^{-1}\cdot \sum_{i=1}^n q_i^j\cdot Y_i(s)\cdot \exp\{\beta_X\cdot x_i+\beta\cdot q_i\}$ ($j\in\mathbb N$). If $Q\noindep U$, we could adjust the effect of $X$ by $Q$ but we could not know the true effect of $Q$. In a randomized clinical trial, both $\beta_Q$ and $\beta_X$ can consistently be estimated through the Cox partial score function and the Breslow estimator.

The second approach adapts an estimation strategy used in observational data analysis to balance two samples with respect to observed values of the covariates by computing a one-dimensional summary of the likelihood that a patient with known characteristics would have been selected in one group versus the other. The resulting summary measure is known as a propensity score. However, the target of inference remains the population-level counterfactual hazard-ratio in which the population is compared counterfactually to itself. That is, it reflects the comparison between the scenarios when all subjects in the population receive one treatment compared to the counterfactual when all subjects receive the other treatment. Based on the PS, we can define a weighting function, $\omega(\cdot)$, which makes that the distribution of the measured confounders invariant between levels of $X$. In this case, the solution of the equation,
\begin{equation}
{\mathscr U}^X_n(\beta,\omega)=\sum_{i=1}^n \int_0^{\infty}\left\{x_i\cdot\omega(q_i) - \frac{{\cal S}^{(1)}_n(X,\omega, \beta,s)}{{\cal S}^{(0)}_n(X,\omega, \beta,s)}\right\}dN_i(s)=0,\label{plw}
\end{equation} 
where ${\cal S}^{(j)}_n(X,\omega, \beta,s)=n^{-1}\cdot \sum_{i=1}^n x_i^j\cdot Y_i(s)\cdot \exp\{\beta\cdot x_i\}\cdot\omega(q_i)$ ($j\in\mathbb N$) yields a consistent estimator of the target parameter. This technique is used, for instance in \citet{todd12}, for controlling survival curves by potential measured confounders, and it is the rationality behind the estimator proposed by \citet{wang18b}.

\section{Estimation of the causal marginal hazard ratio}
\label{iv}
We now consider the case in which the treatment under study is not an exogenous variable, that is, its impact on the failure time is potentially affected by omitted confounders. We assume that we can identify a variable $W$ which satisfies the Assumptions 1-3. Common instruments include prior 
institutional affinity for using a particular procedure, geographic region of residence, an individual's differential access to certain treatments, and an individual's genes aka Mendelian randomization \citep{thanassoulis09}. We consider the estimating equation proposed (without a formal demonstration) by \citet{todd14} for the case of omitted additive confounding. 
\vskip 0.5 cm
\noindent
{\bf Theorem 2.} Let $\{(z_i, \delta_i , x_i, w_i, u_i)\}_{i=1}^n$ be an iid random sample from $(Z,\Delta , X, W, U)$ where the covariate $U$ and the censoring time, $C$, are independent and $W$ satisfies Assumptions 1-3. If the underlying failure time satisfies the Eq. \ref{mcox} model and for each $t>0$, $\int_0^t\mathbb E_X\{{\cal S}^{(2)}_n(W,\beta,s)\}ds<\infty$, then the solution to the function ${\mathscr U}^W_n(\cdot)$, where
\begin{equation}
{\mathscr U}_n^W(\beta)=\sum_{i=1}^n \int_0^{\infty}\left\{w_i - \frac{{\cal S}^{(1)}_n(W,\beta,s)}{{\cal S}^{(0)}_n(W,\beta,s)}\right\}dN_i(s)=0,\label{pl2}
\end{equation} 
and ${\cal S}^{(j)}_n(W,\beta,s)=n^{-1}\cdot \sum_{i=1}^n w_i^j\cdot Y_i(s)\cdot \exp\{\beta\cdot x_i\}$ ($j\in\mathbb N$), is a consistent estimator of $\beta_X$.\par
$\hfill{\Box}$\par
\skip 0.2 cm
\noindent
\phantom b\par
\skip 0.5 cm
\noindent
{\bf Theorem 3.} Under the conditions of Theorem 2, if $\beta^*_n$ is the solution to ${\mathscr U}_n^W(\cdot)=0$, then
\begin{equation}
\sqrt{n}\cdot\frac{ (\beta^*_n - \beta_X)}{\Sigma_n(\beta_X)} \stackrel{\cal L}{\longrightarrow}_n\, {\mathscr N}(0,1), \label{eq:SE}
\end{equation}
where $\Sigma_n^2(\alpha)= \left[\frac{\partial {\mathscr U}_n^W(\alpha)}{\partial\beta}\right]^2\cdot\sum_{i=1}^n \int_0^{\infty}\left\{w_i - \frac{{\cal S}^{(1)}_n(W,\alpha,s)}{{\cal S}^{(0)}_n(W,\alpha,s)}\right\}^2dN_i(s).$\par
$\hfill{\Box}$\par
\phantom b\par
\skip 0.2 cm
In practice, the variance can be approximated by $\Sigma_n^2(\beta^*_n)$. The normal distribution properties guarantees the convergence when we substitute the asymptotic variance with a consistent approximation.\par

\section{Monte Carlo simulation study}
\label{mcs}
Simulating causal scenarios requires randomly generated several inter-related parameters. The random variable modeling the unmeasured confounder, $U$, has to impact both the survival function and the treatment assignment models. For each $i\in\{1,\cdots ,n\}$ ($n=1000$), we run failure times under no treatment  ($X=0$) by generating
\begin{equation}
t_{i,0}=  -\log\{1 - \gamma_{4,1}({u_i} + t_i)\},
\end{equation}
where $u_i$ and $t_i$ are independent random numbers from an exponential (with mean $1$) and a gamma (with parameters 3 and 1) distributions, respectively, and $\gamma_{4,1}(\cdot)$ is the CDF of a gamma distribution with parameters 4 and 1. Then, the failure time for the treated ($X=1$) case is computed as $t_{i,1}= t_{i,0}/\text{HR}_X$ (HR$_X$ stands for the true hazard ratio). The censoring time for the no-treated case, $c_{i,0}$, is generated from an exponential distribution with mean 1 and the censoring time for the treated case, $c_{i,1}$, by $c_{i,0}/\text{HR}_X$ (the expected percentage of censorship is 50\%). The observed time is determined by $z_i=z_{i,0}\cdot I(x_i=0) + z_{i,1}\cdot I(x_i=1)$, where $z_{i,x}=\min(t_{i,x}, c_{i,x})$ and
\begin{equation}
x_i = I(\alpha_U\cdot (u_i -1) + \alpha_W\cdot w_i + \epsilon_i < 0),
\end{equation}
($x_i=0,1$) with $w_i$, $\epsilon_i$ independent standardized normal random variables. Values of $\alpha_U$ and $\alpha_W$ determine the level of endogeneity of the treatment and the strength of the instrument, respectively. Higher values of $\alpha_U$ are directly associated with higher endogeneity but, indirectly, with weaker instruments. Notice that, in the extreme case, the assigned treatment would be absolutely determined by the unmeasured confounders and, therefore, we could not find an instrument variable, $W$, which  simultaneously satisfies Assumptions 2 and 3 \cite{camblor20a}.

\begin{figure}[b]
\centering
\begin{tabular}{ccc}
{$\boldsymbol {\alpha_U=1;\, \alpha_W=1}$} & {$\boldsymbol {\alpha_U=1;\, \alpha_W=2}$}  & {$\boldsymbol {\alpha_U=1;\, \alpha_W=3}$} \\
{\includegraphics[width=4.8 cm]{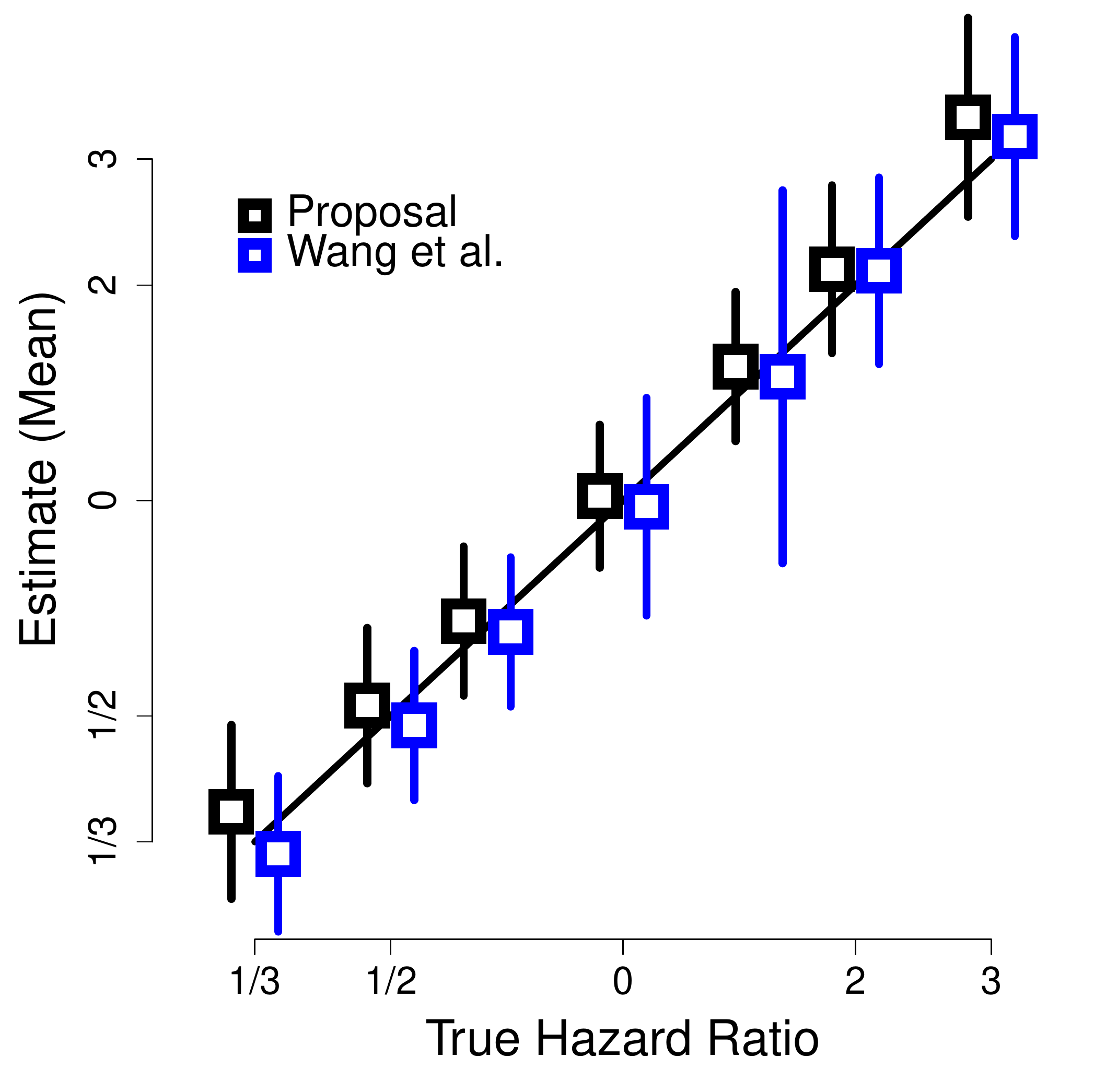}} & \includegraphics[width=4.8 cm]{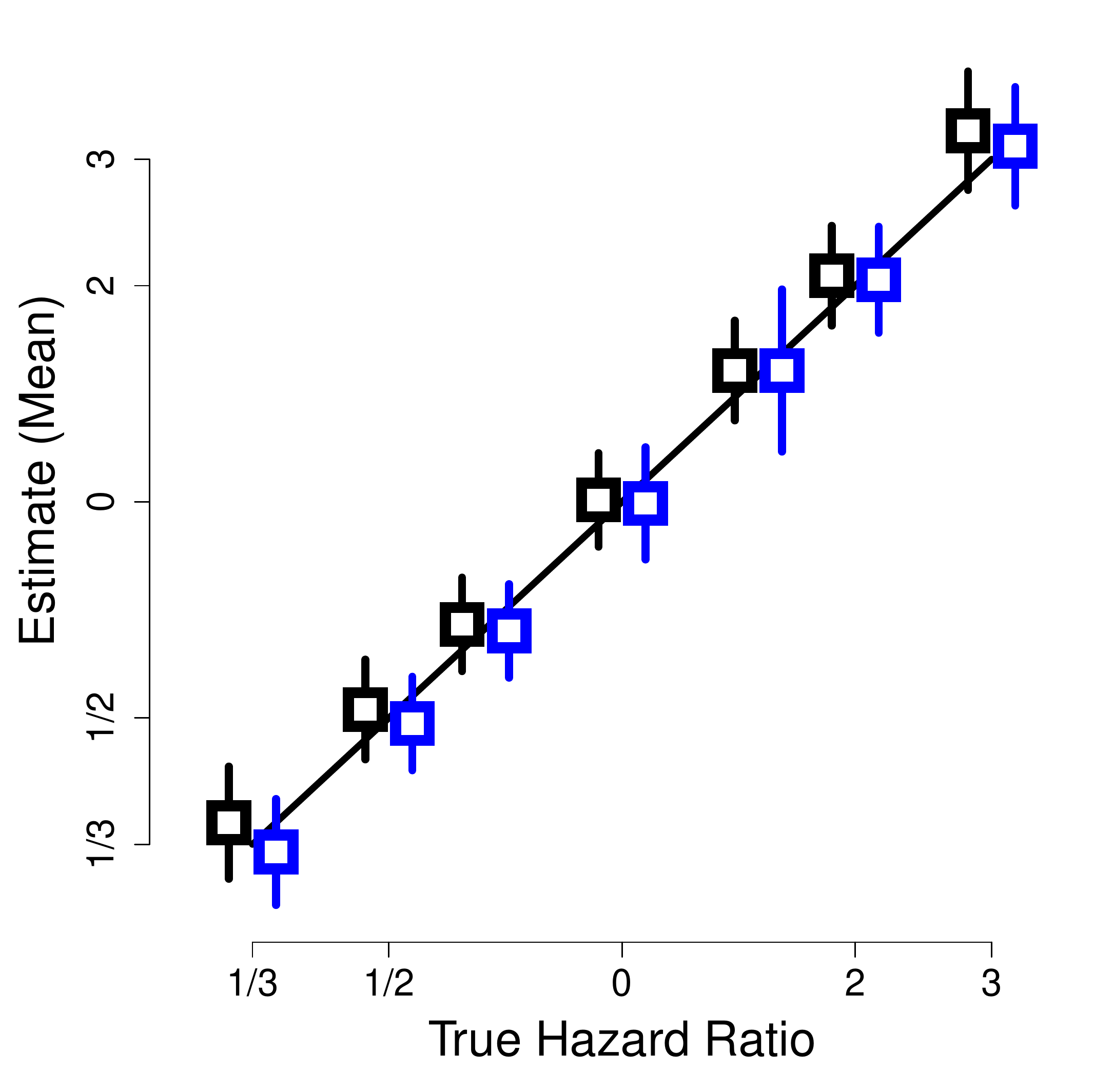} & \includegraphics[width=4.8 cm]{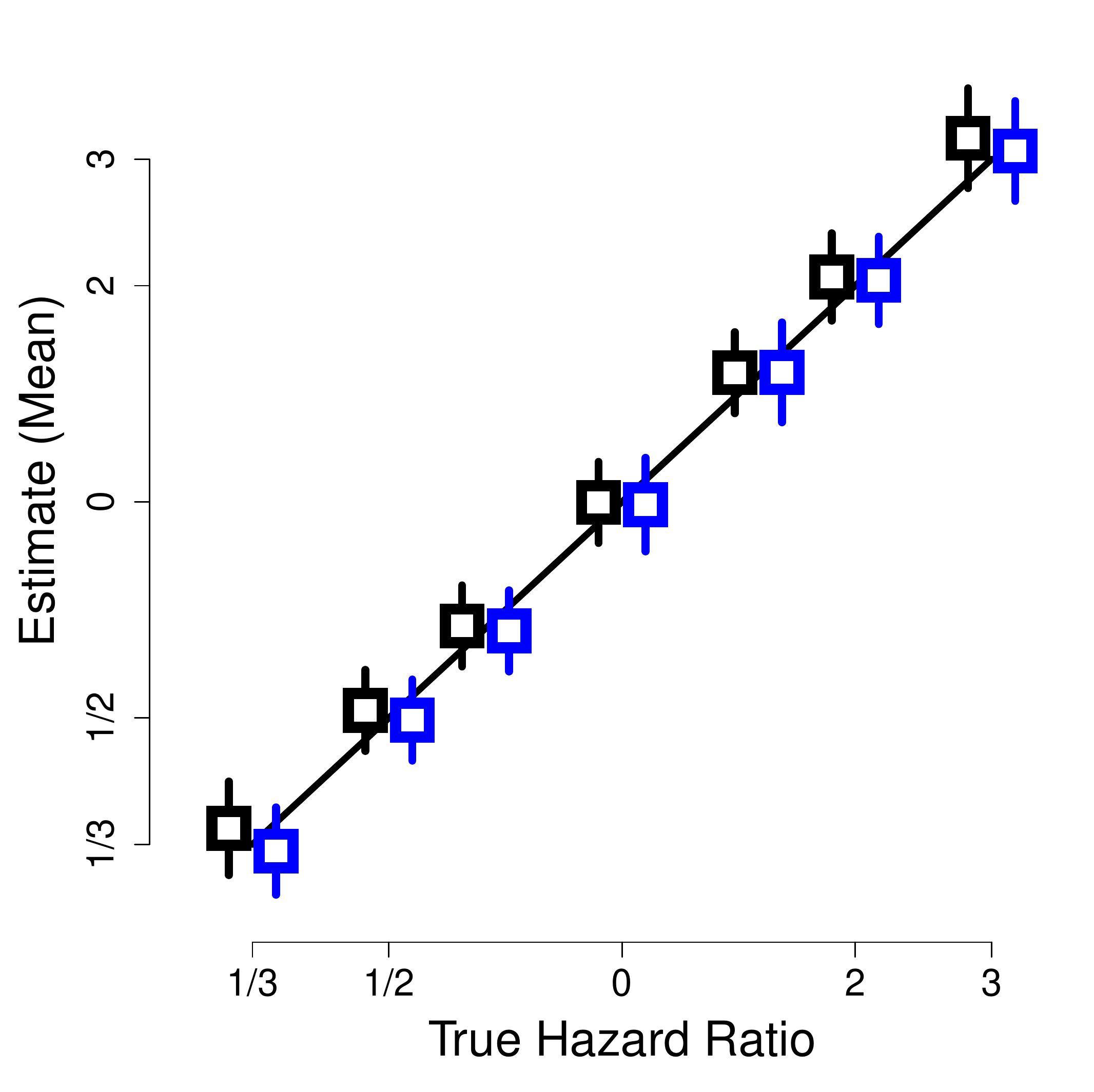}\\
{$\boldsymbol {\alpha_U=2;\, \alpha_W=1}$} & {$\boldsymbol {\alpha_U=2;\, \alpha_W=2}$}  & {$\boldsymbol {\alpha_U=2;\, \alpha_W=3}$} \\
{\includegraphics[width=4.8 cm]{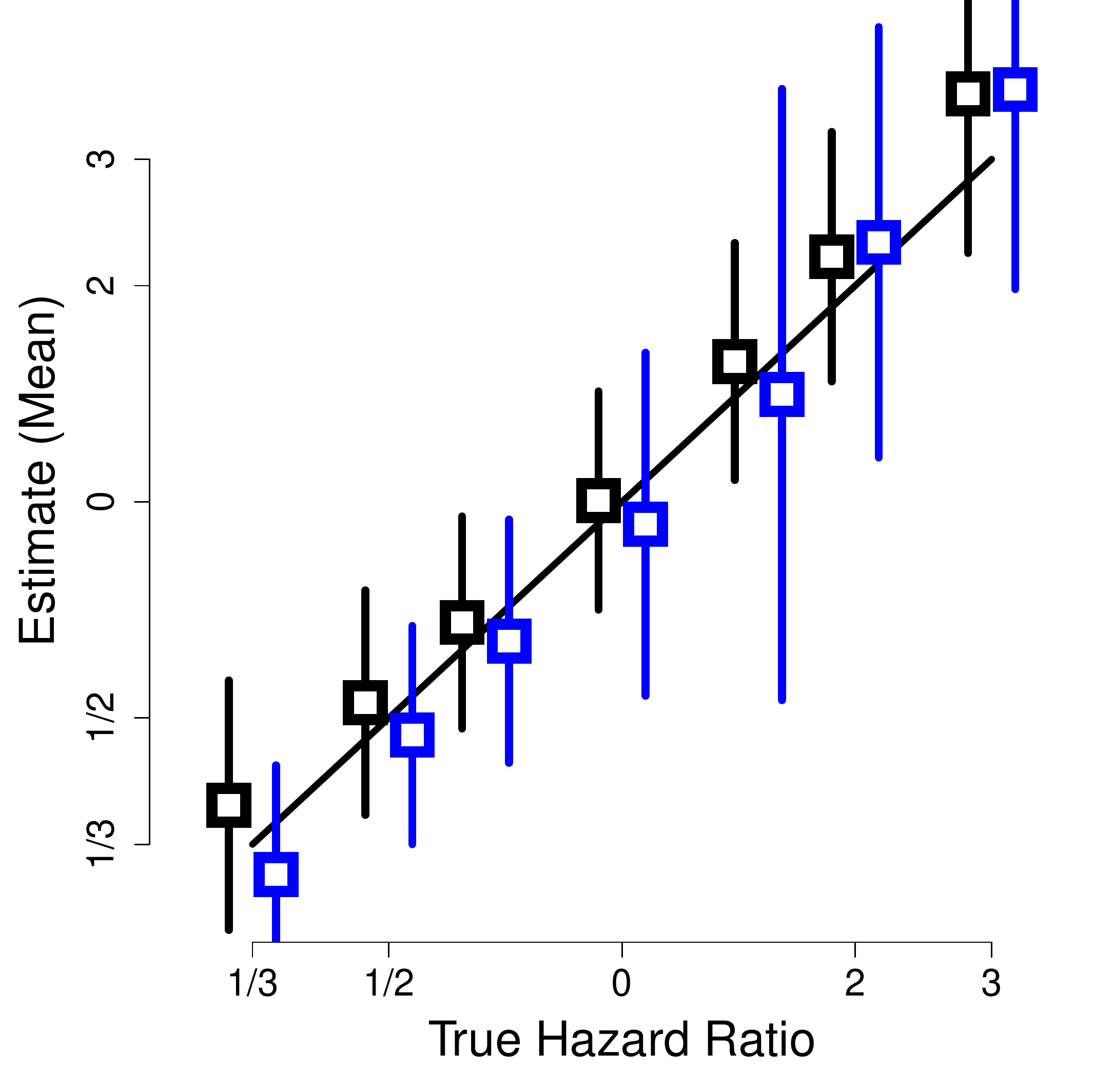}} & \includegraphics[width=4.8 cm]{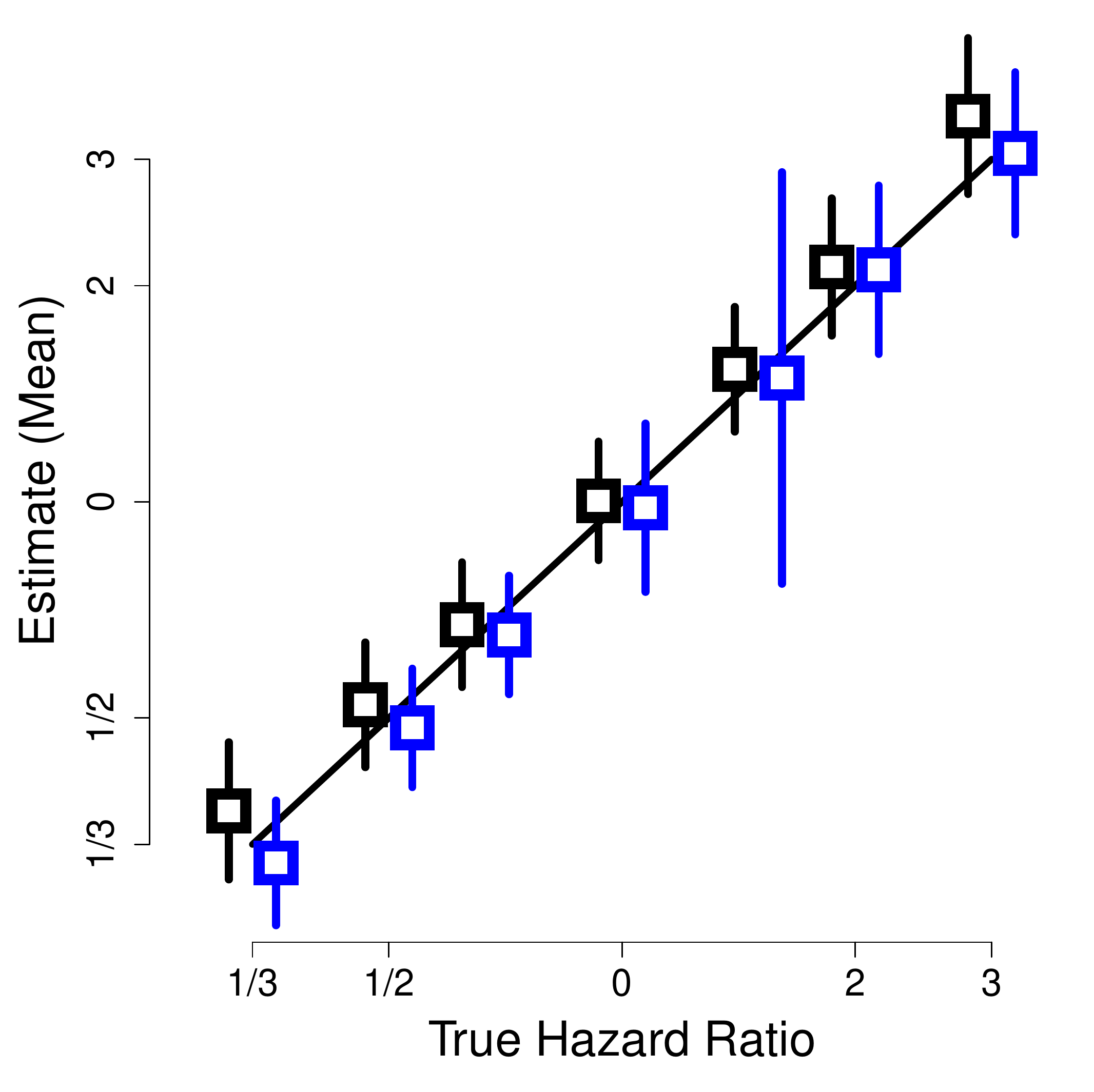} & \includegraphics[width=4.8 cm]{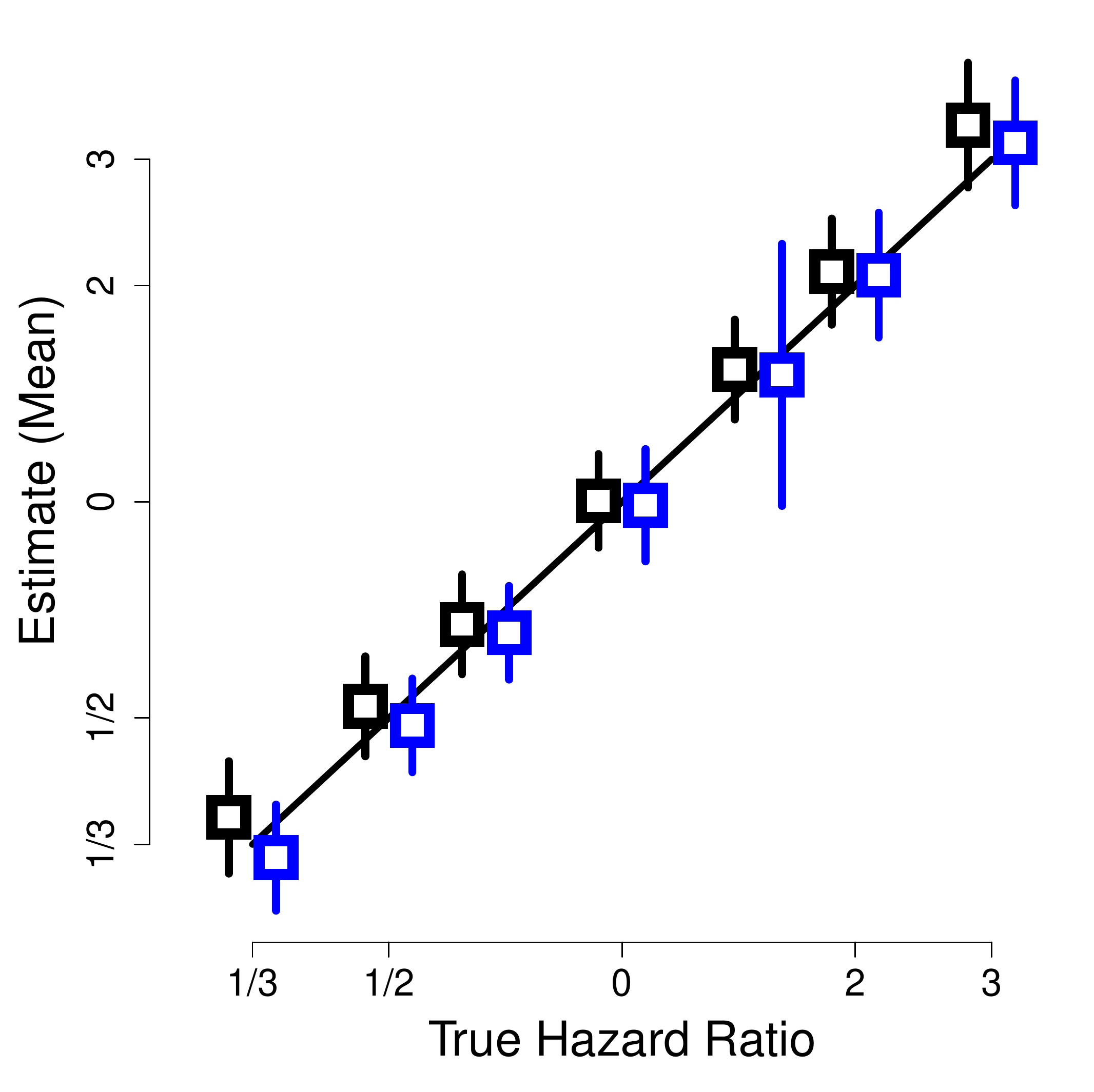}\\
{$\boldsymbol {\alpha_U=3;\, \alpha_W=1}$} & {$\boldsymbol {\alpha_U=3;\, \alpha_W=2}$}  & {$\boldsymbol {\alpha_U=3;\, \alpha_W=3}$} \\
{\includegraphics[width=4.8 cm]{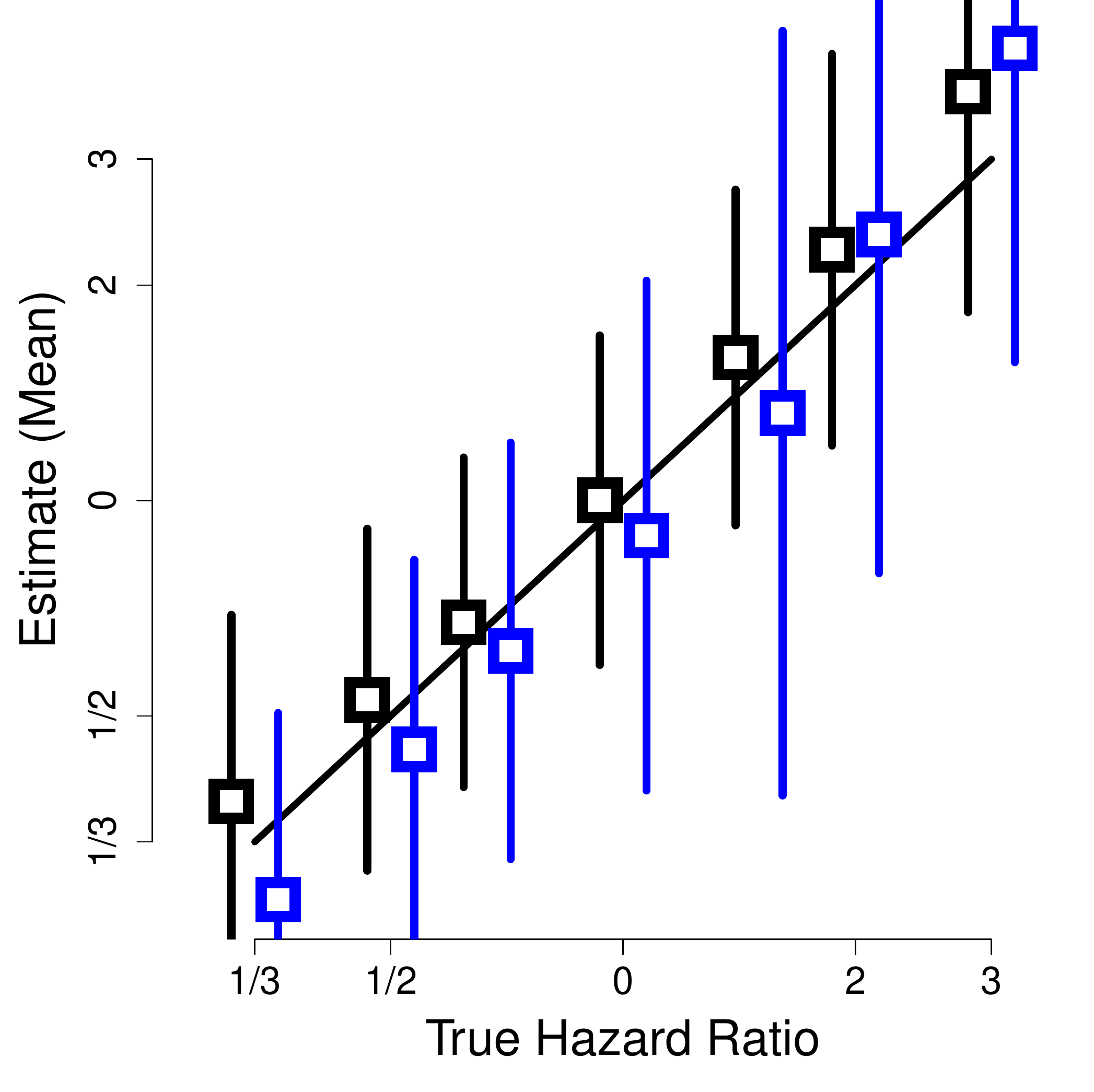}} & \includegraphics[width=4.8 cm]{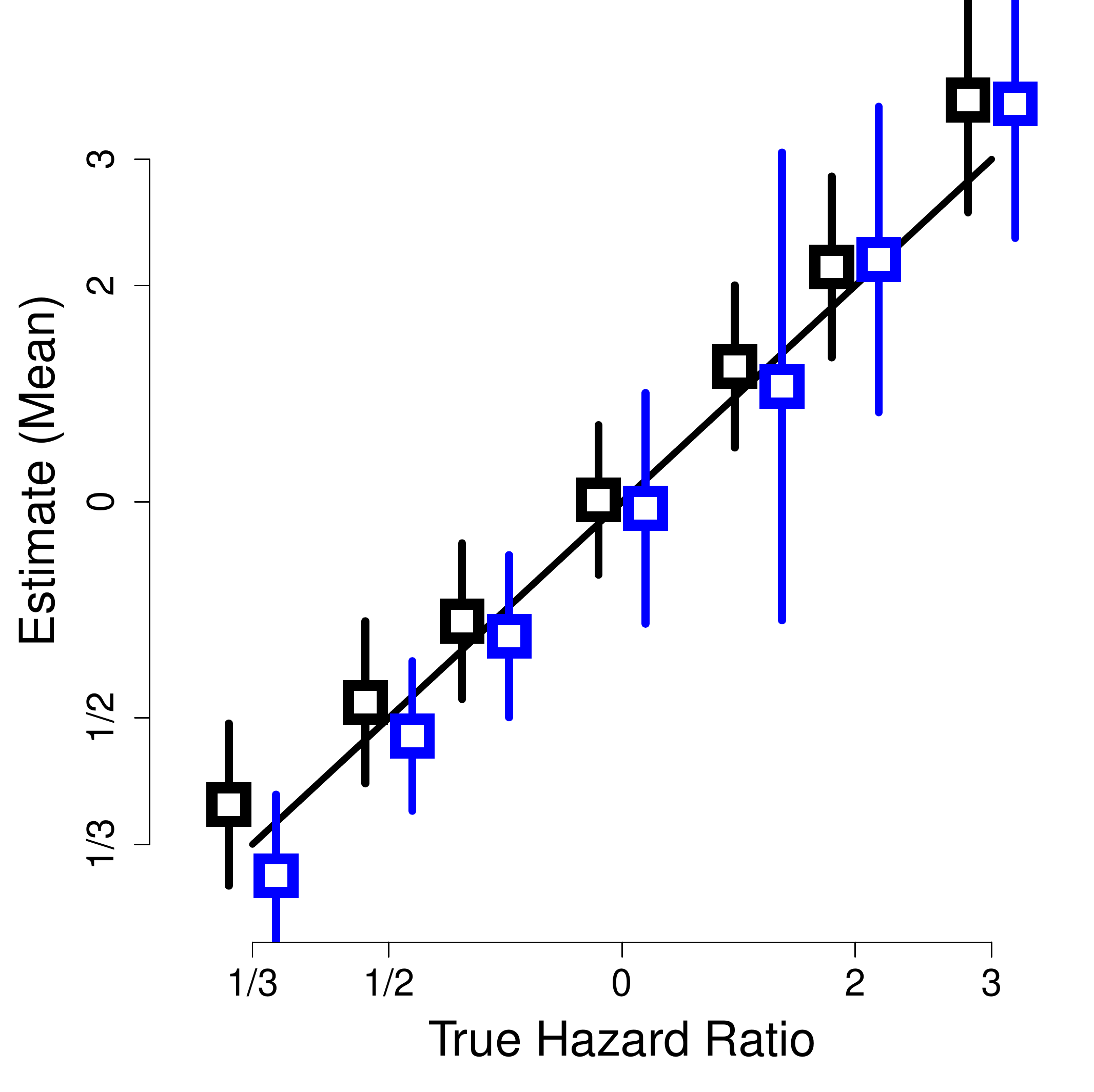} & \includegraphics[width=4.8 cm]{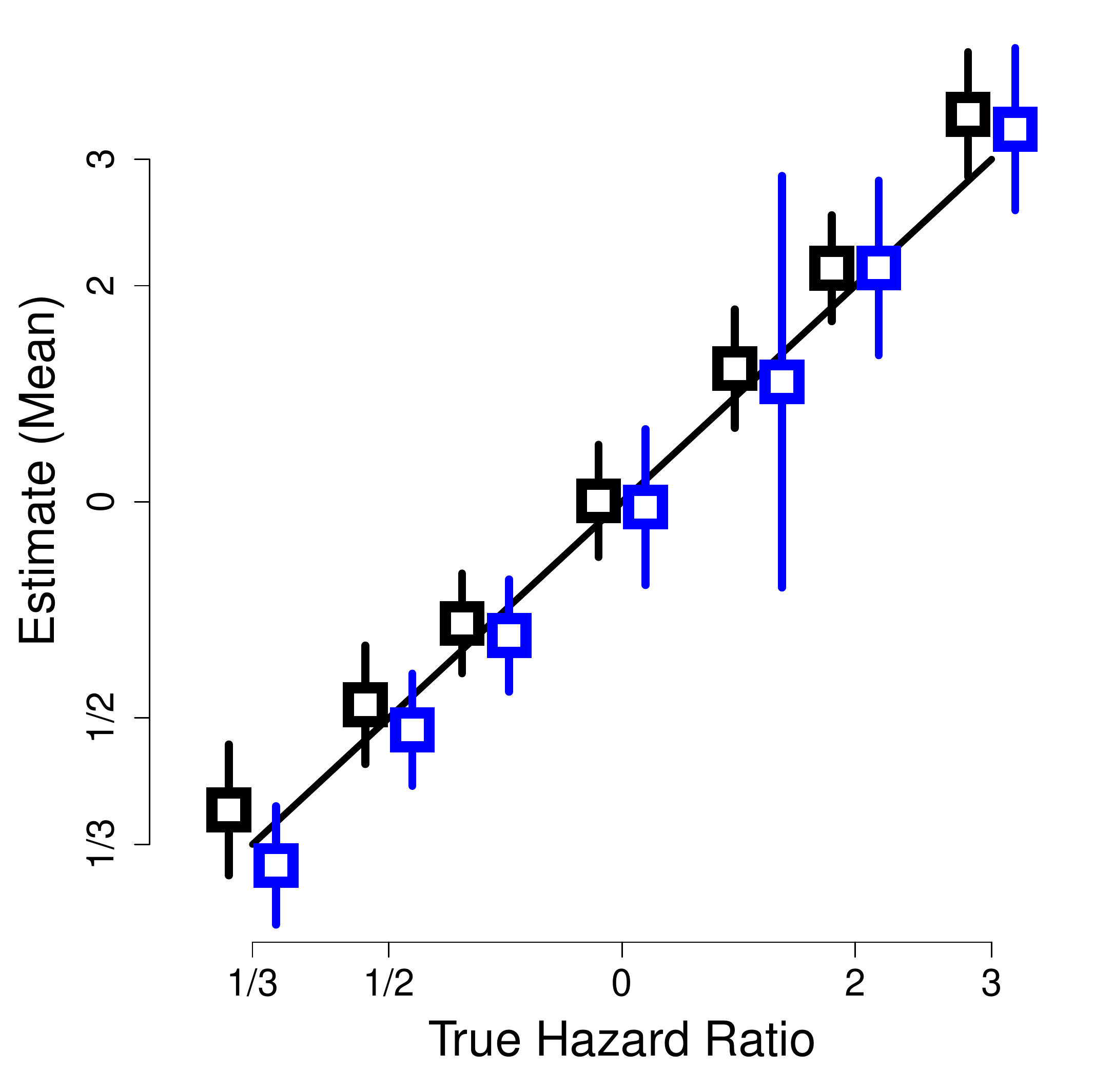}\\
\phantom b\par
\end{tabular}
\caption{Means (squares) $\pm$ standard deviations (vertical lines) for the ICC model (black) and the \citet{wang18b} (blue) estimators for 5000 Monte Carlo iterations of the models described in Section 5.
\label{simul}}
\end{figure}

Figure \ref{simul} shows the mean and the standard deviation observed in 5000 Monte Carlo iterations for different values of $\alpha_U$, $\alpha_W$ and HR$_X$ for both the proposed and the \cite{wang18b} estimators of $\beta_X$ ($\log(\text{HR}_X)$). For the Wang et al.  estimator (see the Appendix for more details about this estimator), standard deviations were approximated through 50 bootstrap \cite{efron93} iterations. Both estimators report similar results. The Wang et al. estimator reports slightly more biased results in most of the configurations (in particular, in 33 out of 63). It obtained better results for the extreme values of HR$_X$. However, in some of the considered situations, it failed to report a numeric solution; for instance, for $\alpha_U=3$, $\alpha_W=1$ and HR$_X=3/2$, in 950 out of 5000 iterations (19\%) it did not reach a numerical solution (the IIC model estimator failed in 10 out 5000 iterations, 0.2\%). The Wang et al. estimator was also sensitive to the IV's strength; for $\alpha_U=3$, it is the winner in only 7 out 21 configurations. Both procedures provided adequate 95\% confidence intervals. Coverage percentages of our new estimator ranged between 92.8\% and 97.3\% while the bootstrap standard deviation for the Wang et al. estimator yields slightly more conservative confidence interval (coverage percentages ranged between 94.8\% and 98.5\%).

\section{Practical application}
\label{example}
With the purpose of illustrating the practical implementation of the proposed methodology, we study the effect of the treatment used for carotid revascularization (carotid endarterectomy versus transfemoral carotid artery stenting) among patients with carotid artery stenosis using data from the vascular quality initiative, VQI, registry.

The VQI (\url{www.vascularqualityinitiative.org}) is a nationwide registry which collects perioperative and one-year follow-up data to generate real-time benchmarked reports to assess the quality of care and determine best practices in vascular surgery. We are interested in determining the effect of the type of intervention received for carotid revascularization among patients with carotid artery stenosis: transfemoral carotid endarterectomy stenting (CAS) versus transcarotid artery revascularization (TCAR), on time to event (death by any cause, any perioperative stroke or long-term ipsolateral stroke). Although TCAR has been used as an alternative technique for patients at high surgical risk for endarterectomy, several studies have shown that CAS has a higher periprocedural stroke risk compared with carotid endarterectomy (see \cite{schermerhorn20} and references therein). 

For an instrumental variable (IV) we consider the preference at the level the medical center for using TCAR the prior three months. Patient specific values of IV are given by the proportion of TCAR performed in the same hospital over the 3 months prior to their procedure adjusted for patient characteristics, calendar day of surgery, and total number of carotid surgeries over the previous three months. The time-varying proportion of past cases is justified as an instrument due to: 1) hospitals that performed a high relative amount of a certain procedure in the past are likely to keep doing so; 2) there should be no effects of the relative frequency of therapies performed on a patient's outcome except through its impact on treatment choice for that patient; 3) we know of no factors that would directly influence a hospital's performance conditional on adjusting for the total number of procedures performed at the center over the past 3 months; hence, our inclusion of the total number of procedures as a covariate in our analysis.

\begin{figure}[b]
\centering
\includegraphics[width=15 cm]{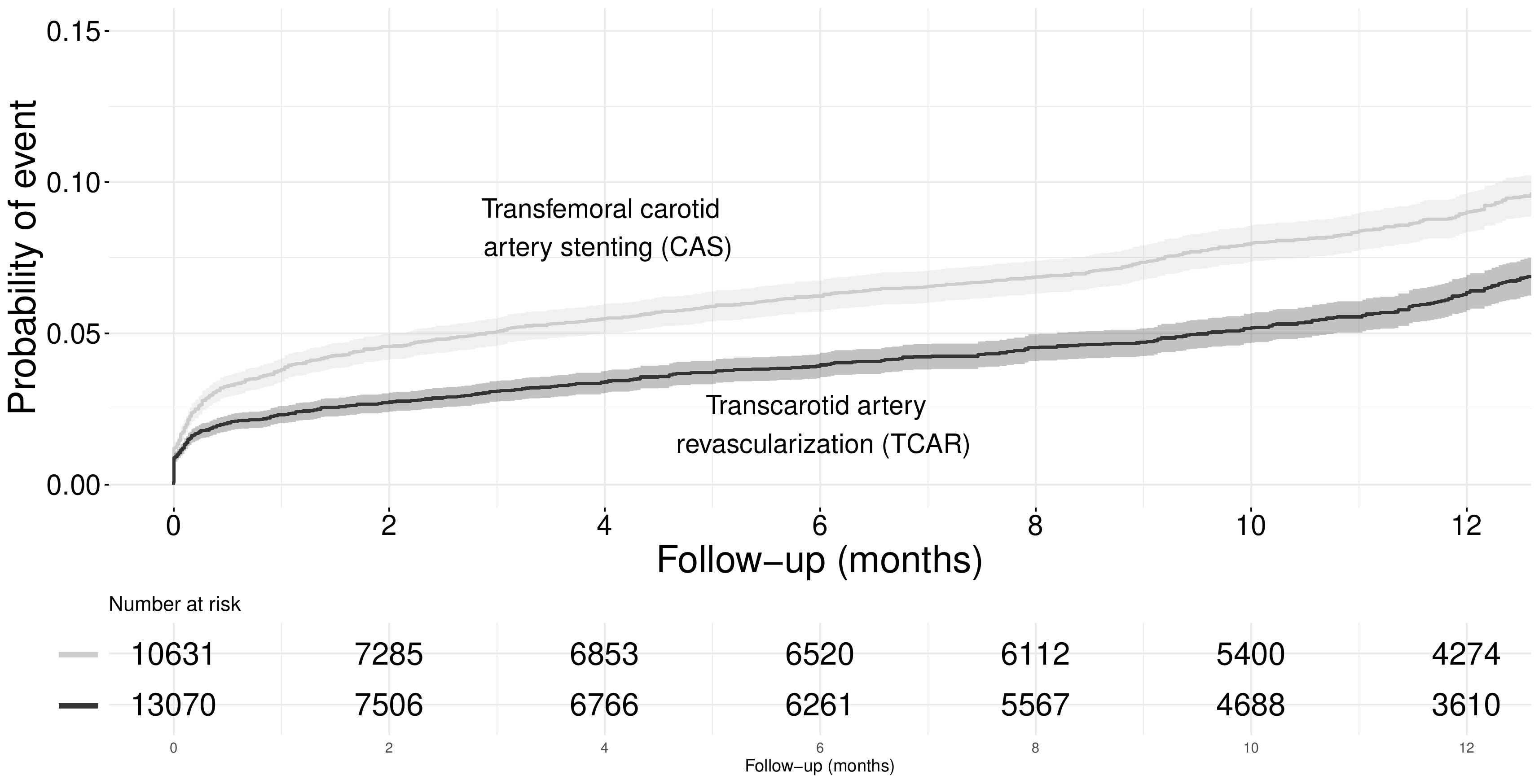}\\
\phantom b\par
\phantom b\par
\phantom b\par
\caption{Kaplan-Meier estimators for the time to event in CAS and TCAR patients.\label{survival}}
\end{figure}

A total of 10631 patients receiving CAS and 13071 patients receiving TCAR from January 2016 to May 2020 were included. The total number of events recorded were 942 and 640 with incidence rates per 100 patient years of 11.04 and 8.32 for CAS and TCAR patients, respectively. Figure \ref{survival} represents the time to event distribution in both the CAS and TCAR groups and the number of patients at risk during the follow-up time.

The crude (standard) hazard ratio was 0.71 (95\% confidence interval; 0.63 to 0.79). When we adjust by potential (measured) confounders, the HR was 0.75 (0.64 to 0.89). The considered instrumental variable has a strong relationship with the treatment (F-statistic of 4,946.3). The Cohen's Kappa index for the resulting dichotomization (based on the Youden index) was 0.83 (0.82 to 0.84); 91.8\% of patients were allocated in the actual received therapy. Figure \ref{iv} depicts the density (left-panel) of the IV (expit transformed) and the respective violin plot (right-panel) by treatment group. For the Wang et al. estimator, which only works with binary IVs, we used the dichotomized IV and the naive bootstrap algorithm with 500 iterations for computing the standard deviation

\begin{figure}[b]
\centering
\begin{tabular}{cc}
\includegraphics[width=7 cm]{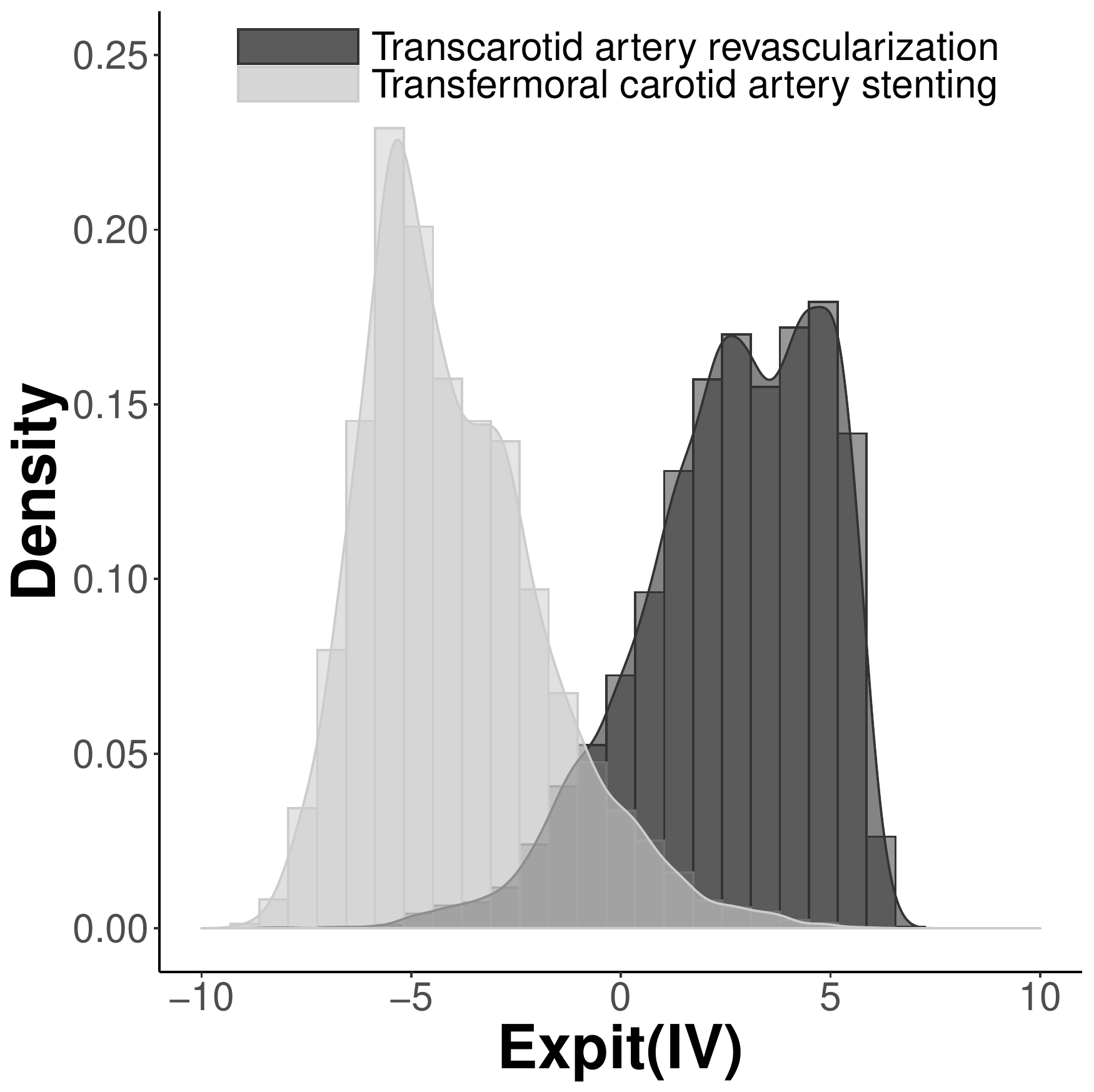} & \includegraphics[width=7 cm]{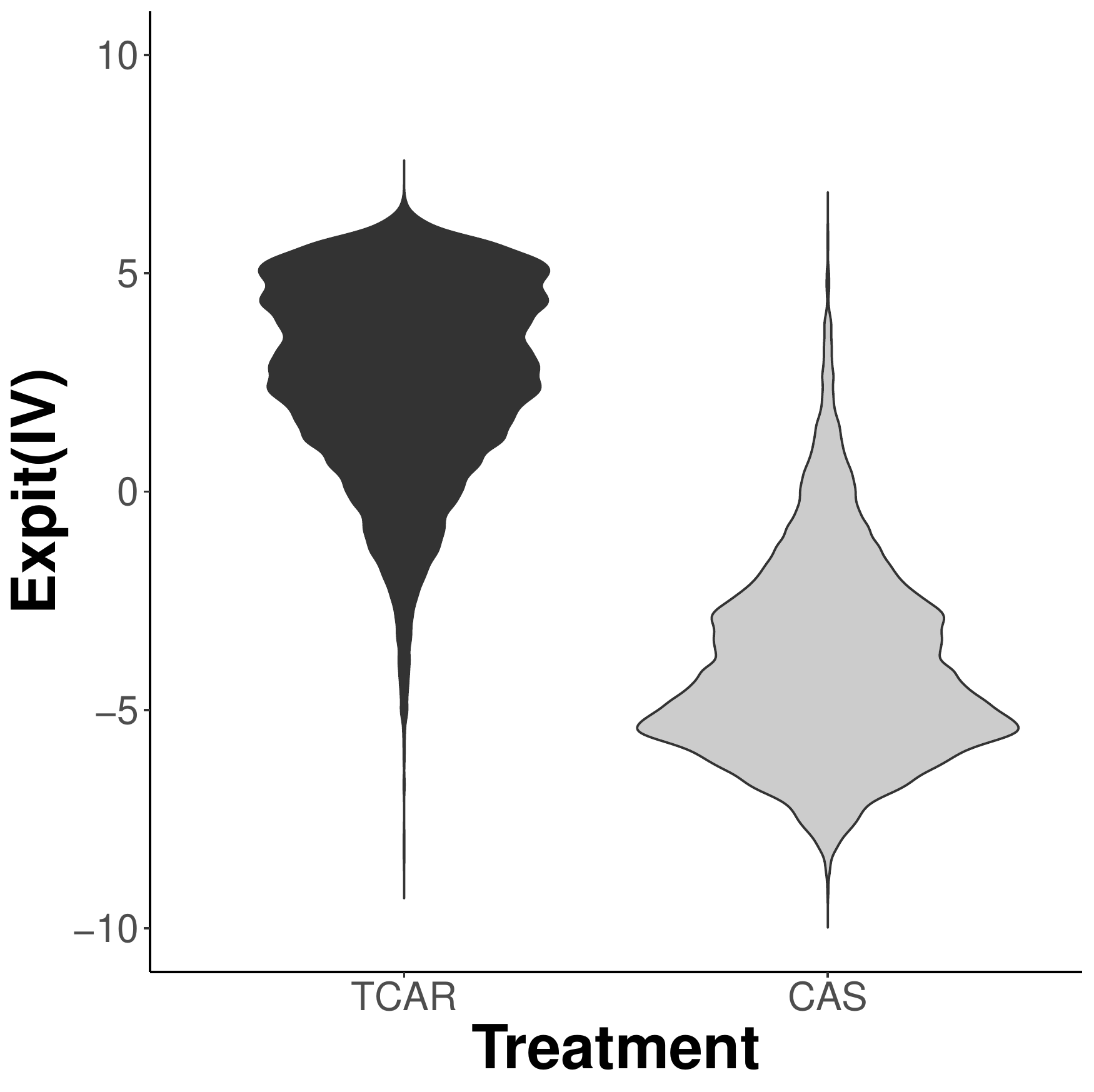}\\
\phantom b\par
\phantom b\par
\phantom b\par
\end{tabular}
\caption{Density (left) and violin (right) plots for the expit transformation of the instrumental variable by treatment group.\label{iv}}
\end{figure}

\cite{wang18b} reported a populational hazard ratio of 0.67 (0.37 to 1.23) while the results reported by our proposed procedure were even more favorable to TCAR with a populational hazard ratio of 0.63 (0.56 to 0.71). The propensity score inverse weighting (PSIW) estimations for the hazard ratios from the standard Cox regression were 0.62 (0.59 to 0.64) and 0.61 (0.55 to 0.67) for the crude and adjusted models, respectively. For Wang et al., it was 0.68 (0.38 to 1.24) and, for the IIC model it was 0.69 (0.61 to 0.78). Not surprisingly, since we adjusted for potential confounders, the results of the marginal Cox regression were similar in both unweighted and weighted populations. Both Wang et al. and the proposed procedure reported similar results, consistent with the results of our simulation study in the case of moderate effect and strong instrumental variable. Notice that the Wang el al. estimator also reported large variability in several of the settings in the Monte Carlo simulations, likely caused by its dependency of the IV's strength. The difference with the standard Cox regression hazard ratios suggests  the presence of omitted confounders.

\section{Discussion}
The Cox regression model and the associated hazard ratio are enormously popular for summarizing the effect of treatments on
  time to event outcomes. Notwithstanding this, the hazard ratio's interpretation strongly depends on the assumed underlying survival model and its use has been criticized recently. The impact of omitted covariates on the resulting hazard ratio, even in absence of an interaction between these covariates and the treatment, and the potential effect of those on the profile of people at risk after the initiation of the study, introduces concerns about the hazard ratio interpretation \citep{hernan10, aalen15, martinussen13, martinussen20}.   

Several alternative indices have been proposed in order to deal with these concerns \citep{todd14, martinussen20, camblor20}. The marginal Cox regression models include an identifiable hazard ratio with populational causal interpretation. The marginal Cox regression model does not allow adjustment for an unmeasured confounder. That is, the hazard ratio is the result of comparing each subject in the studied population among all of the same subjects when the level of the treatment would counterfactually be different. In this sense, we cannot interpret those results as the individual counterfactual but as the population counterfactual. 

We studied the properties of an instrumental variable procedure for its estimation in the presence of omitted confounder (see \citet{tchentgen15} for a complete discussion of the instrumental variable procedures in the survival context). We prove the large sample consistency of the estimation score proposed by \citet{todd14}. Our Monte Carlo simulation study suggests that finite sample behavior of the procedure is adequate in the considered scenarios. It is more robust than its competitors for weak instruments, although it is more biased for large effects of the treatment. Notice that, for excessively weak instruments, both procedures could fail to obtain a solution since the estimation equation could became numerically unsoluable. In the Monte Carlo simulations, this ocurred more frequently in the Wang et al. procedure than in the new procedure proposed herein. Beyond that, the new procedure is more flexible than Wang's since it works for non-binary treatments and non-binary instruments.

The results obtained in the practical examples are consistent with our Monte Carlo simulations. With a strong instrument and a moderate effect of the treatment reports similar results for both the Wang et al. and our new estimator. Differences with standard Cox regression models suggest the presence of omitted confounders.

The procedure developed in this paper, may be generalized to the case when multiple instruments are available. One possible means of proceeding is to form an equation emulating Equation (\ref{pl2}) for each available instrument, $w_{1}, ..., w_{M}$, where $M$ is the number of instruments. The construction of these equations reflects that each IV generates a separate equation that can be solved. Let $\hat{\beta}_{m}$ be the solution of the $m$-th equation and let $s(\hat {\beta}_{m})$ be its associated standard error ($1\leq m \leq M$) . A natural pooled estimator that combines the constrained maximum-partial-likelihood (cMPLE) estimators for each IV individually into a single estimator is then
\begin{equation}
 \hat{\beta} = \frac{\sum_{m=1}^{M}  s(\hat{\beta}_{m})^{-1} \hat{\beta}_{m} }{ \sum_{m=1}^{M} s(\hat{\beta}_{m})^{-1} }. \label{eq:MultIV}
 \end{equation}
The resulting estimator may be thought of as a pooled cMPLE. An alternative way of combining the estimators would be to solve the sum (or weighted sum) of the instrument-specific expressions in Equation (\ref{pl2}) over the respective equations. If the impact of each IV is proportional to $s(\hat{\beta}_{m})^{-1}$ then the resulting estimator will be asymptotically equivalent to the quantity in (\ref{eq:MultIV}). This procedure holds irrespective of the number of exogenous  covariates.

\section*{Supplementary Material}
The routines used for implementing (in R) the studied procedure are provided.

\section*{Acknowledgements}
The authors are grateful with Prof. Linbo Wang for sharing his code with us and with Dr. Jesse Columbo and Phillip Goodney for providing the data for real-world example.

\section*{Conflict of interest}
The authors have no conflicts of interest to report.

\section*{Appendix}
\subsection*{Results' proof}
\noindent
{\it Proof of Theorem 1.}

Under the stated assumptions, we know (\citet{truthers86}) that the solution to ${\mathscr U}^X_n(\cdot)$ is a consistent estimator for the solution to
\begin{align*}
{\mathscr U}^X_T(\beta)=& \int {\mathbb E}_{X,U}\left\{ x\cdot\lambda_x(t,u)\cdot S_x(t,u)\cdot G_x(t)\right\}\cdot {\mathbb E}_{X,U}\{ e^{\beta\cdot x}\cdot S_x(t,u)\cdot G_x(t)\}dt\\
&- \int{\mathbb E}_{X,U}\left\{ \lambda_x(t,u)\cdot S_x(t,u)\cdot G_x(t)\right\}\cdot {\mathbb E}_{X,U}\{ x\cdot e^{\beta\cdot x}\cdot S_x(t,u)\cdot G_x(t)\}dt,
\end{align*}
where $G_x(\cdot)={\mathscr P}\{ C>t|X=x\}$. From the Eq. \ref{mcox} (and the Fubini's Theorem) we have
\begin{align*}
\frac{\partial}{\partial t}\log\left( {\mathbb E}_U\{e^{-\Lambda_x(t;u)}\}\right)= \exp\{\beta_X\cdot x\}\cdot \log\left( {\mathbb E}_U\{e^{-\Lambda_0(t;u)}\}\right),
\end{align*}
and therefore
\begin{align*}
{\mathbb E}_U\{\lambda_x(t;u)\cdot S_x(t,u)\}=& \exp\{\beta_X\cdot x\}\cdot \frac{ {\mathbb E}_U\{\lambda_0(t;u)\cdot S_0(t,u)\}}{ {\mathbb E}_U\{S_0(t,u)\}}\cdot {\mathbb E}_U\{ S_x(t,u)\}\\
=&\kappa_U(t)\cdot e^{\beta_X\cdot x}\cdot {\mathbb E}_U\{ S_x(t,u)\}.
\end{align*}
Then the independence between $X$ and $U$ implies
\begin{align*}
{\mathscr U}^X_T(\beta)=& \int \kappa_U(t)\cdot {\mathbb E}_X\left\{x \cdot e^{\beta_X\cdot x}\cdot G_x(t)\cdot {\mathbb E}_U\{ S_x(t,u)\}\right\}\cdot {\mathbb E}_X\left\{e^{\beta\cdot x}\cdot G_x(t)\cdot {\mathbb E}_U\{ S_x(t,u)\}\right\}dt\\
& -\int \kappa_U(t)\cdot {\mathbb E}_X\left\{ e^{\beta_X\cdot x}\cdot G_x(t)\cdot {\mathbb E}_U\{ S_x(t,u)\}\right\}\cdot {\mathbb E}_X\left\{x \cdot e^{\beta\cdot x}\cdot G_x(t)\cdot {\mathbb E}_U\{ S_x(t,u)\}\right\}dt,
\end{align*}
which has a unique solution at $\beta=\beta_X$.\par
$\hfill{\Box}$\par
\vskip 0.5 cm
\noindent
{\it Proof of Theorem 2.}

From Assumption 1 ($W\indep T|X$) we have that, for $\beta_W=0$, the true survival function satisfies
\begin{align}
S_{x,w}(t)=&{\mathbb E}_U\left\{{\mathscr P}\{T>t|X=x, W=w, U=u\} \right\}\nonumber \\ 
=&{\mathbb E}_U\left\{{\mathscr P}\{T>t|X=0, W=0, U=u\} \right\}^{e^{\beta_X\cdot x + \beta_W\cdot w} }= S_{0,0}(t)^{e^{\beta_X\cdot x+ \beta_W\cdot w}}. \label{aux}
\end{align}
The maximum partial-likelihood estimator of the parameter ${\boldsymbol \beta}_X=(\beta_X, \beta_W)$ is based on the maximization of the function,
\begin{align*}
\ell ({\boldsymbol \beta})=\sum_{i=1}^n \int \log\{\lambda_{x_i}&(t,u_i,w_i; {\boldsymbol\beta})\cdot Y_i(t)\}dN_i(t) \nonumber\\
- \int \log&\left\{\sum_{i=1}^n\mathbb E_U\{\lambda_{x_i}(t,u,w; {\boldsymbol\beta})\cdot Y_i(t)\}\right\}d\sum_{i=1}^n N_i(t),
\end{align*}
where ${\boldsymbol\beta}=(\beta_1,\beta_2)$. Then, $\boldsymbol\beta_X$ is a solution to the partial derivative equation of ${\mathbb E}_{X,W}\{\ell ({\boldsymbol \beta})\}$. From Eq. \ref{aux} and the Assumption 2 ($W\indep U|X$), we have that $\boldsymbol\beta_X$ is a solution for
\begin{align*}
0=&{\mathbb E}_{X,W}\left\{\frac{\partial \ell ({\boldsymbol \beta})}{\partial \beta_2}\right\}\\
=&{\mathbb E}_{X,W}\left\{\sum_{i=1}^n \int_0^{\infty}\left\{w_i - \frac{ \sum_{i=1}^n w_i\cdot Y_i(s)\cdot \exp\{\beta_1\cdot x_i + \beta_2\cdot w_i\}}{ \sum_{i=1}^n Y_i(s)\cdot \exp\{\beta\cdot x_i+\beta_2\cdot w_i\}}\right\}dN_i(s)\right\}
\end{align*}

Assumption 1 ($W\indep T|X$) guarantees that $\beta_W=0$ and therefore ${\mathbb E}_{W,X}\{{\mathscr U}_n^W(\beta_X)\}=0$. In addition, we have that
\begin{small}
\begin{align*}
{\mathbb E}_{X,W}\left\{\frac{\partial {\mathscr U}_n^W(\beta)}{\partial \beta}\right\}=&{\mathbb E}_{X,W}\left\{\int_0^1\frac{ \sum_{i=1}^n x_i\cdot Y_i(s)\cdot \exp\{\beta\cdot x_i\}\cdot \sum_{i=1}^n w_i\cdot Y_i(s)\cdot \exp\{\beta\cdot x_i\}}{\left(\sum_{i=1}^n Y_i(s)\cdot \exp\{\beta\cdot x_i\}\right)^2}d\sum_{i=1}^nN_i(s) \right\}\\
& -{\mathbb E}_{X,W}\left\{\int_0^1 \frac{ \sum_{i=1}^n w_i\cdot x_i\cdot Y_i(s)\cdot \exp\{\beta\cdot x_i\}\cdot \sum_{i=1}^n Y_i(s)\cdot \exp\{\beta\cdot x_i\}}{\left(\sum_{i=1}^n Y_i(s)\cdot \exp\{\beta\cdot x_i\}\right)^2}d\sum_{i=1}^n N_i(s) \right\}\\
=& {\mathbb E}_{X,W}\left\{\int_0^1 \frac{ \sum_{j=1}^n\sum_{i=1}^n (x_i\cdot w_j - x_i\cdot w_i)\cdot Y_i(s) Y_j(s)\cdot \exp\{\beta\cdot (x_i+x_j)\}}{\left(\sum_{i=1}^n Y_i(s)\cdot \exp\{\beta\cdot x_i\}\right)^2}d\sum_{i=1}^n N_i(s) \right\}
\end{align*}
\end{small}
The Cauchy-Schwartz  inequality and Assumption 3 ($W\noindep X$) guarantee that this is a non-zero function with constant sign and hence, ${\mathbb E}_{W,X}\{{\mathscr U}_n^W(\cdot)\}$ has one unique zero reached at $\beta_X$.\par
$\hfill{\Box}$\par
\vskip 0.5 cm
\noindent
{\it Proof of Theorem 3.}

Asymptotic normality of $\beta_X$ is directly derived from M-statistics theory (see, for instance, \citet{van00}). From the Theorem 2 and the Taylor expansion, we have that
\begin{equation*}
\sqrt{n}\cdot (\beta^*_n - \beta_X)=\frac{-\sqrt{n}\cdot{\mathscr U}_n^W(\beta_X)}{\frac{\partial {\mathscr U}_n^W(\beta_X)}{\partial\beta}+\frac{1}{2} \frac{\partial^2 {\mathscr U}_n^W(\bar \beta_n)}{\partial\beta^2}(\beta^*_n - \beta_X)},
\end{equation*}
where $\bar\beta_n$ is a point between $\beta_X$ and $\beta_n^*$. From Theorem 2, the central limit theorem and the Slutsky lemma, we have that $\sqrt{n}\cdot{\mathscr U}_n^W(\beta_X)$ is asymptotically normal with mean zero and variance
\begin{equation*}
{\mathbb V}\{\sqrt{n}\cdot{\mathscr U}_n^W(\beta_X)\}=\sum_{i=1}^n \int_0^{\infty}\left\{w_i - \frac{{\cal S}^{(1)}_n(W,\beta_X,s)}{{\cal S}^{(0)}_n(W,\beta_X,s)}\right\}^2dN_i(s).
\end{equation*}
Theorem 2 also implies that $(\beta_n^* - \beta_X)=o_{\mathscr P}(1)$. Therefore, the variance of $\sqrt{n}\cdot (\beta^*_n - \beta_X)$ is
\begin{equation*}
{\mathbb V}\{\sqrt{n}\cdot (\beta^*_n - \beta_X)\}= \left\{ \frac{\partial {\mathscr U}_n^W(\beta_X)}{\partial\beta}\right\}^{-2}\cdot{\mathbb V}\{\sqrt{n}\cdot{\mathscr U}_n^W(\beta_X)\},
\end{equation*}
and the proof is concluded.\par
$\hfill{\Box}$\par
\subsection*{Wang et al. estimator}
Let $\{(x_i, z_i, \delta_i, q_i, w_i)\}_{i=1}^n$ be an iid random sample containing the treatment, the observed event time, the observed status (failure versus censoring), the measured covariates and the instrumental variable (now assumed to be binary), respectively. Wang et al. \cite{wang18b} propose to estimate $\beta$ by solvin for $\beta$ the equation
\begin{equation*}
\sum_{i=1}^n \delta_i\cdot \hat\omega (w_i, q_i)\cdot \left[x_i - \frac{\sum_{j=1}^n x_j\cdot e^{\beta\cdot x_j}I(z_j\geq z_i)\cdot\hat\omega (w_i, q_i) }{\sum_{j=1}^n e^{\beta\cdot x_j}I(z_j\geq z_i)\cdot\hat\omega (w_i, q_i)} \right],
\end{equation*} 
where $\hat\omega (w_i, q_i)= h(x_i)\cdot (2w_i -1)/\{f(w_i|q_i: \hat\eta)\cdot \delta^X(q_i; \hat\gamma)$, with $h(\cdot)$ any function of $X$ such that the above equation is well-defined. There are different procedures for the estimation of the parameters of the density, $f(W|Q)$, and the conditional risk difference,  $\delta^X(Q; \hat\gamma)$, functions. We refer to \citet{wang18b} for specific details about the procedure.

\bibliographystyle{unsrtnat}
\bibliography{Bibliography.tex}
\end{document}